\documentclass[twocolumn,superscriptaddress]{revtex4}
\usepackage{graphicx}

\begin{document}

\title{ COMPLEXITY IN STRONGLY CORRELATED \\
ELECTRONIC SYSTEMS}

\author{Elbio Dagotto}
\affiliation{Department of Physics, University of Tennessee, 
                   Knoxville, Tennessee 37996-1200,
                   and Condensed Matter Sciences Division, Oak Ridge National Laboratory,
                   Oak Ridge, Tennessee 37831-6393, USA}

\maketitle

\section{ABSTRACT}

{\bf 
A wide variety of experimental results and theoretical investigations
in recent years 
have convincingly demonstrated that several transition metal oxides
and other materials, 
have dominant states that are not spatially homogeneous. This 
occurs in cases in which several 
physical interactions -- spin, charge, lattice, and/or orbital --
are simultaneously active. This phenomenon causes 
interesting effects, such as colossal magnetoresistance, 
and it also appears crucial
to understand the high temperature 
superconductors. The spontaneous emergence of electronic
nanometer-scale structures in transition metal oxides, and the existence of
many competing states, are properties 
often associated with complex matter where nonlinearities dominate,
such as soft materials and biological systems. 
This electronic complexity could have potential consequences 
for applications of correlated electronic materials, because
not only charge (semiconducting electronic), or charge and spin (spintronics)
are of relevance, but in addition the lattice and orbital degrees
of freedom are
active, leading to giant responses to small perturbations. 
Moreover, several metallic and insulating
phases compete, increasing the potential for novel behavior.
}

\vskip 0.5cm

Materials in which the electrons are strongly correlated
display a broad range
of  interesting phenomena, including 
colossal magnetoresistance (CMR) where enormous variations in
resistance are produced by small magnetic field changes,
and high temperature superconductivity (HTSC).
An important characteristic
of these materials is the existence of several competing states, as exemplified
by the complicated phase diagrams that  transition metal oxides (TMOs)
present (Fig.1).
The understanding of these oxides 
has dramatically challenged our view of solids. 
In fact, after one of the largest 
research efforts ever in physics, involving hundreds of scientists, 
even basic properties of
the HTSC cuprates, such as
the pairing mechanism, linear resistivity, and pseudogap phase, are still only poorly understood.
In the early days of HTSC, it was expected that suitably
modified theories of ordinary metals would explain the unusual properties
of the cuprate's normal state.
However, important experimental results gathered in recent years
have revealed an unexpected property of oxides:
Many TMOs are  inhomogeneous at the nanoscale (and sometimes  
at even longer length scales).
This explains why the
early theories based on homogeneous systems were not successful, and
raises hopes that a novel avenue for progress
has opened.

What are the implications of these and other results reviewed below?
It will be argued that
the current status of correlated electrons investigations must be considered in
 the broader context of complexity.
In his pioneering article \cite{anderson}, 
Anderson wrote that ``the ability to
reduce everything to simple fundamental laws does not imply the ability to start
from those laws and reconstruct the universe''. 
In complex systems \cite{complexity}, the properties of a few
particles are not sufficient to understand large aggregates 
when these particles strongly interact. Rather, in such systems, which are not
merely complicated, one expects 
emergence, namely the generation of properties that do not preexist 
in a system's constituents.
This concept is contrary to the philosophy of reductionism,
the traditional physics hallmark. 
Complex systems spontaneously tend to form structures (self-organization), 
and these structures vary widely in size and scales.
Exceptional events are important, as when the
last metallic link completes a 
percolative network. The average 
behavior is of no relevance for this phenomenon, and often only a few rare
events dominate. Evidence is accumulating, that TMOs and related materials
have properties similar to standard complex systems, and several
results must be reexamined in this broader framework.

\begin{figure}
\includegraphics[width=4.7cm]{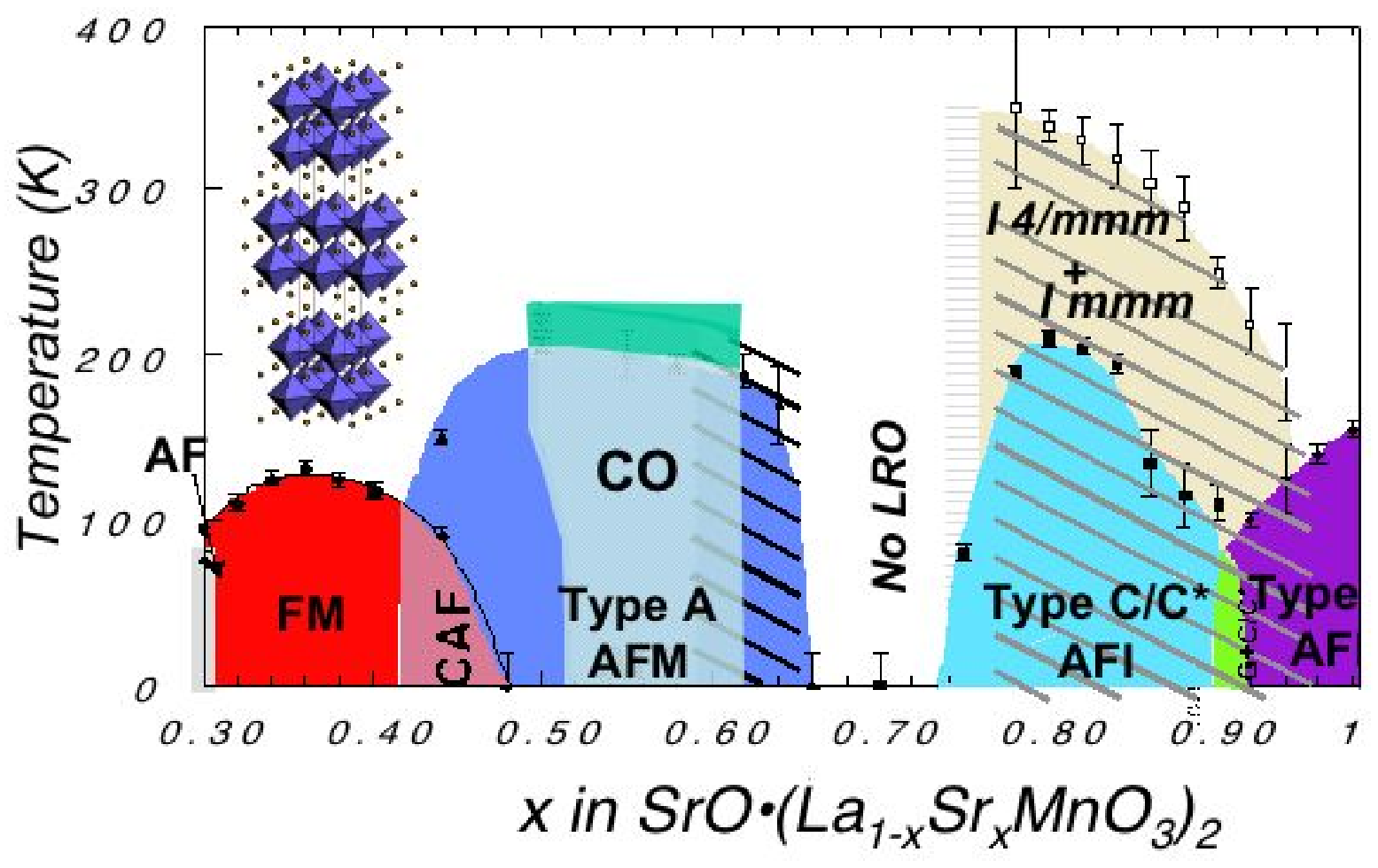}
\includegraphics[width=3.3cm]{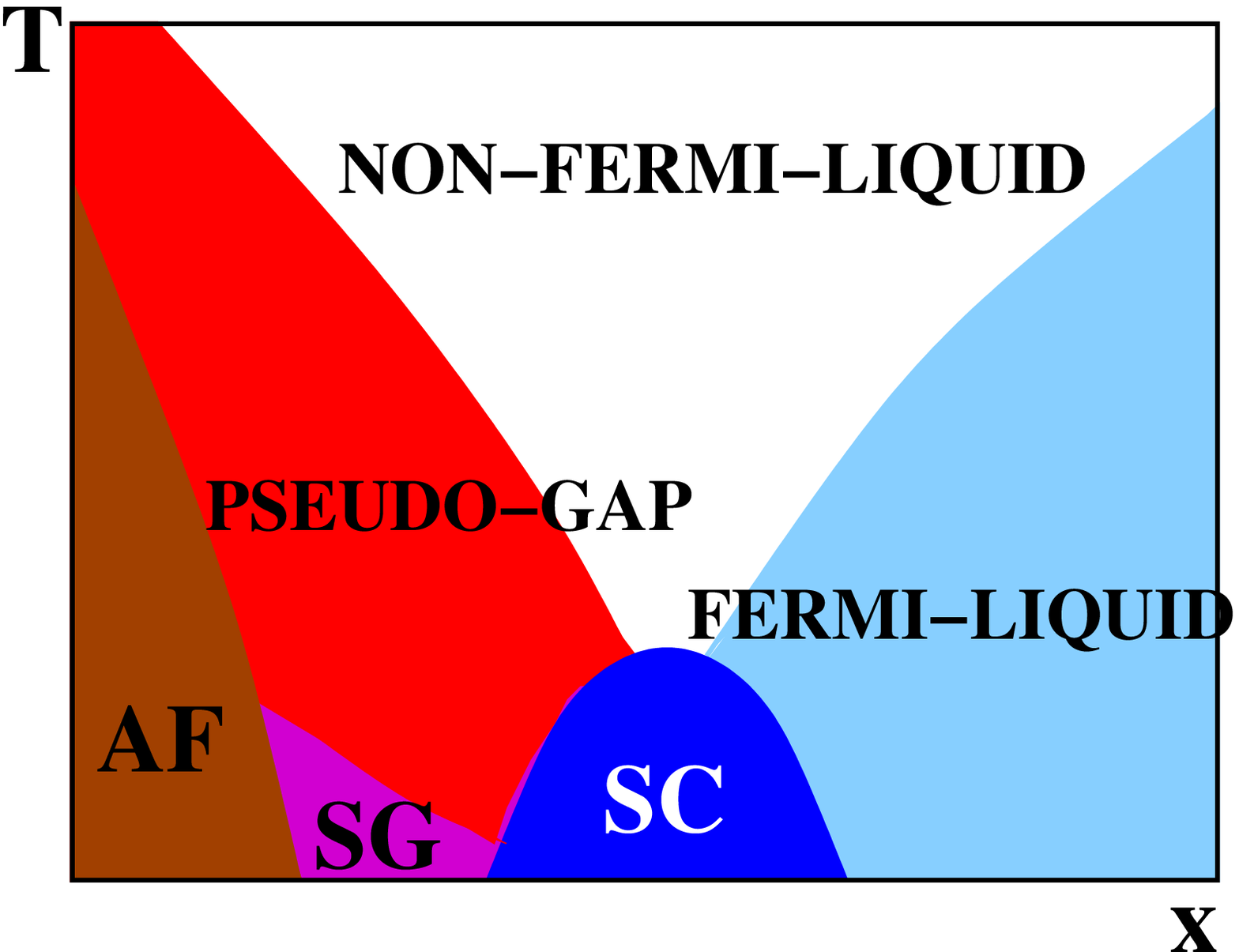}
\includegraphics[width=4cm,clip]{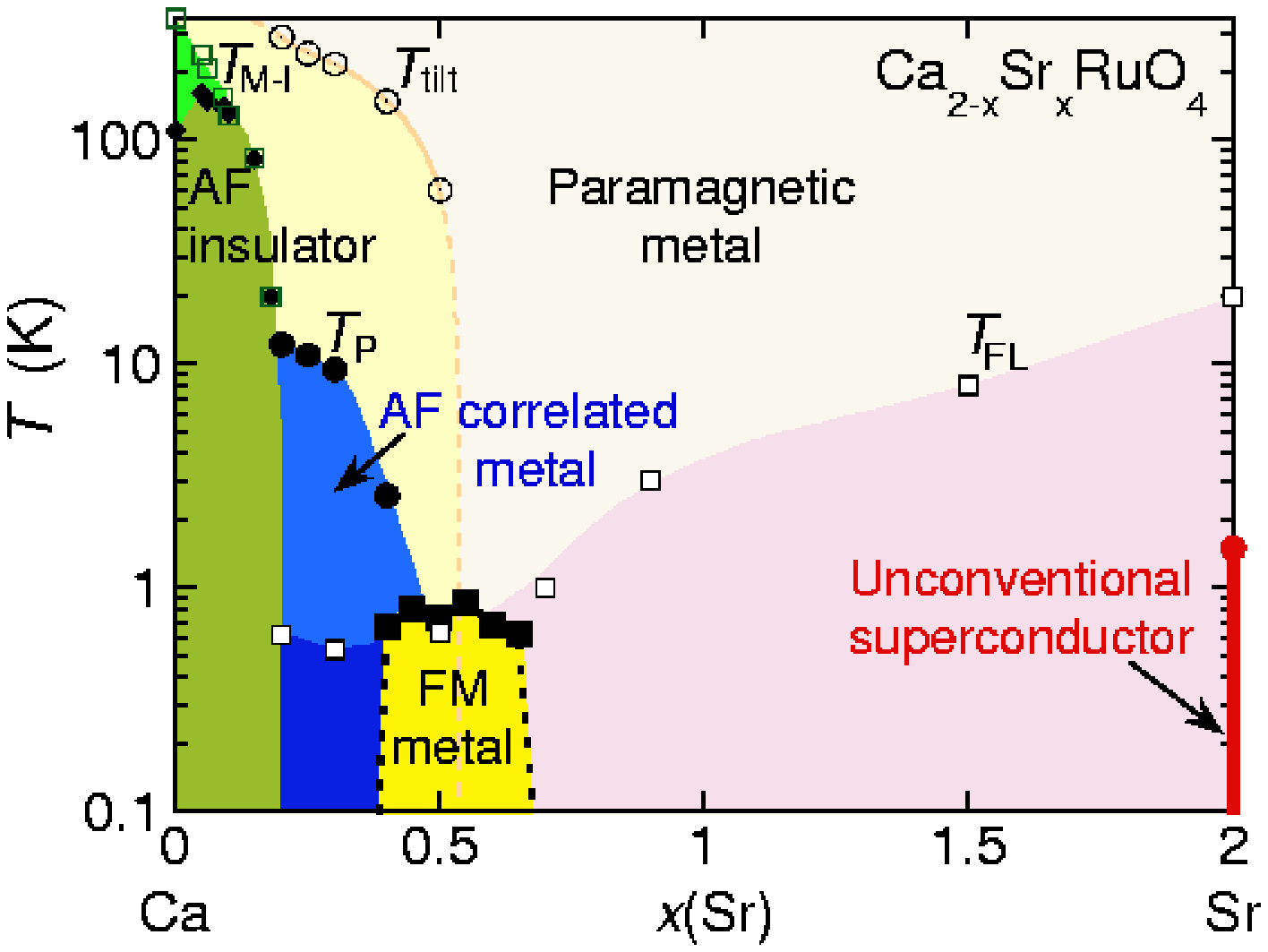}
\includegraphics[width=4cm]{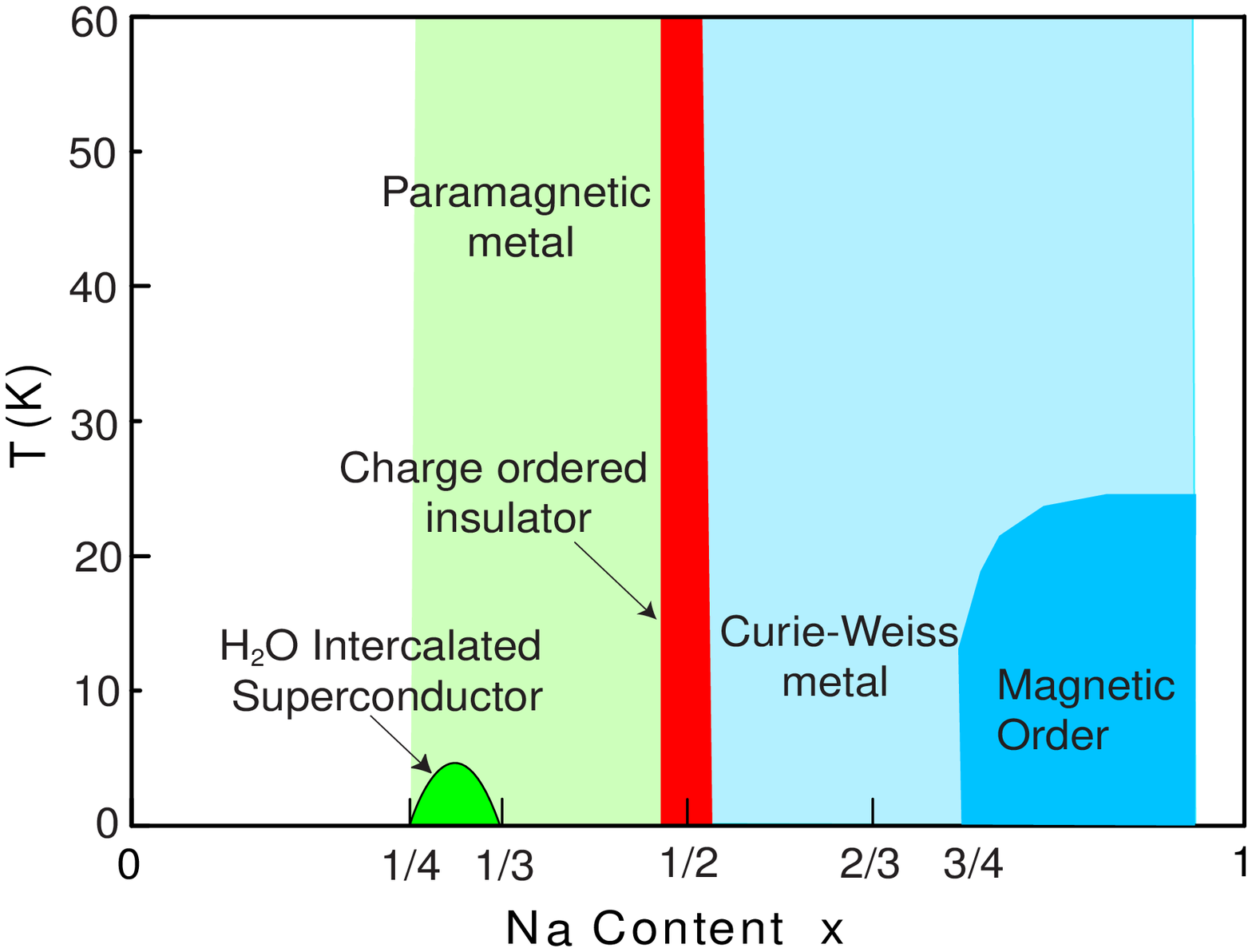}
\includegraphics[width=4cm]{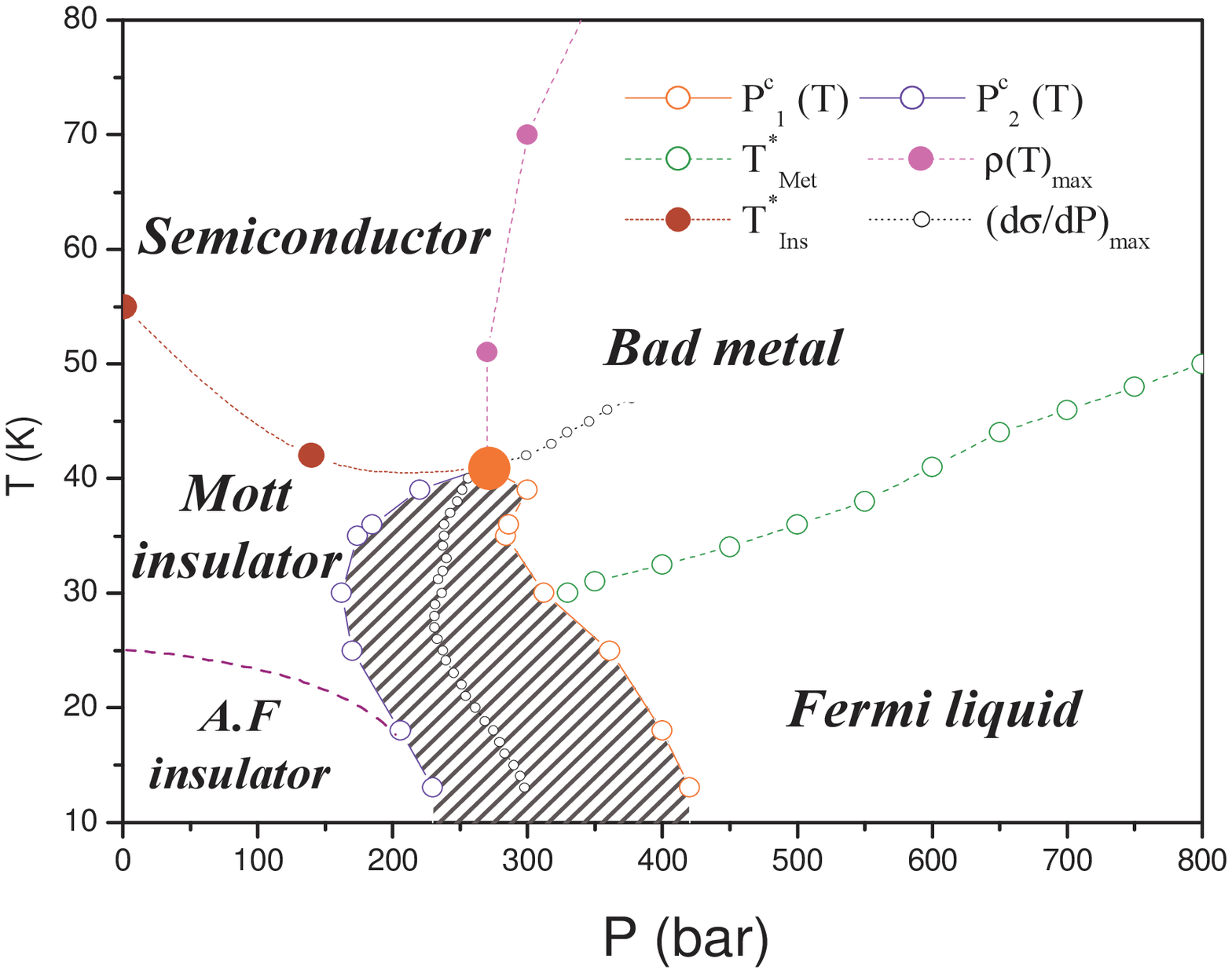}
\includegraphics[width=4cm]{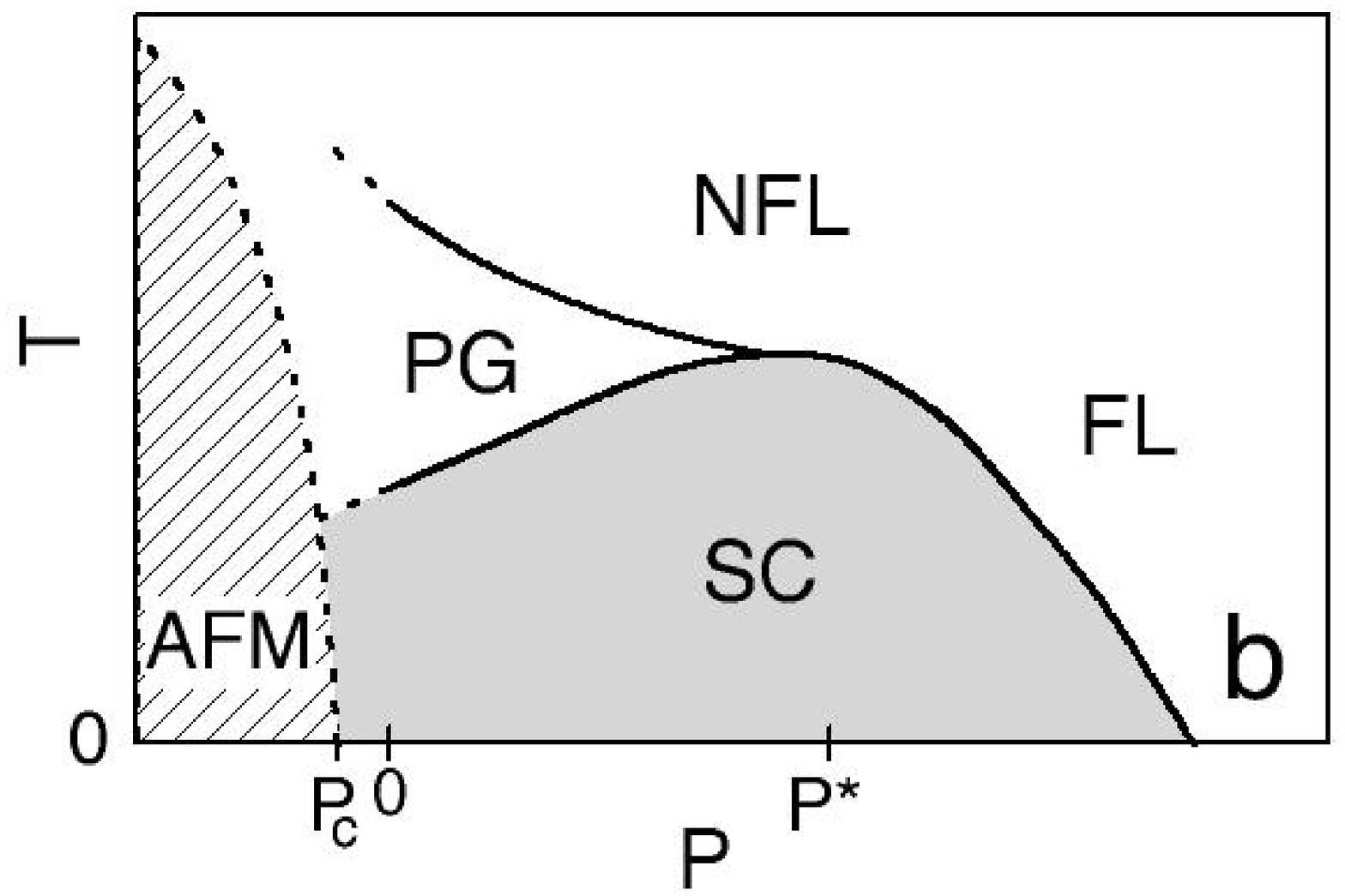}
\caption{\label{fig:phase-diagrams} 
Phase diagrams of representative materials of the 
strongly correlated electron family (notations are standard and
details can be found in the original references).
(a) Temperature versus hole density
phase diagram of bilayer manganites \cite{mitchell}, including several
types of antiferromagnetic (AF)  phases, a ferromagnetic (FM) phase, and even
a globally disordered region at $x$=0.75.
(b) Generic phase diagram for HTSC.
SG stands for spin glass.
(c) Phase diagram of single layered ruthenates \cite{nakatsuji,nakatsuji2}, 
evolving from a superconducting 
(SC) state at $x$=2 to an AF insulator at $x$=0
($x$ controls the bandwidth rather than
the carrier density). Ruthenates are believed to be clean metals at
least at large $x$, thus providing a family of oxides 
where competition and complexity can
be studied with less quenched disorder than in Mn-oxides.
(d) Phase diagram of Co oxides
 \cite{cava}, with SC, 
charge-ordered (CO), and magnetic regimes.
(e) Phase diagram of the organic
$\kappa$-(BEDT-TTF)$_2$Cu[N(CN)$_2$]Cl salt \cite{organic}.
The hatched region denotes the coexistence of metal and insulator phases.
(f) Schematic phase diagram of the 
Ce-based heavy fermion materials \cite{heavy}.
} 
\end{figure}

\vskip 0.3cm
\noindent {\bf Nanostructures in Manganites and Cuprates}
\vskip 0.2cm

\noindent {\it Manganites:} 
The Mn oxides called manganites \cite{adriana,review,book,tokura,salamon,mathur,ahn2},
especially those displaying the CMR effect,
are an important oxide family in which the presence of inhomogeneous states
is widely accepted. A remarkable 
cross-fertilization between theory and experiments 
has led to considerable progress in unraveling the 
role of these inhomogeneities.
Theoretical investigations \cite{review} 
predicted that, in a broad region 
of parameter space, the ground state is actually a nanoscale
mixture of phases, particularly 
in the presence of quenched disorder \cite{burgy,burgy2,yunoki}, 
namely, when random ``frozen''
deviations from
the perfectly uniform system are incorporated in the study. Many 
experimental results are indeed
in agreement with the basic notion that the relevant phases
are not homogeneous; these results also  provide information crucial
to understanding the CMR effect \cite{review,book,cheong,egami}.
Some of the general theoretical ideas 
are summarized in the schematic phase diagram (Fig.2a) \cite{burgy}, which has been experimentally 
confirmed \cite{tokura-disorder,tokura-disorder2} (Fig.2b).

In the clean limit without quenched disorder,
the two key competing states in manganites, ferromagnetic (FM)
metallic  and antiferromagnetic (AF) insulating (AFI), are known to be
separated by a first-order transition \cite{review,book}. However,
once the inevitable quenched disorder is included in
the calculation, arising, for example, 
from the lattice-distorting chemical doping procedure, 
non-statistical fluctuations of dopant density, or strain fields,
the region in which the two states are nearly degenerate 
(that is, they can coexist) is
dramatically modified. 
In this regime, there is still a local 
tendency toward either FM or AFI short-distance correlations. 
However, globally, 
neither of the two states dominates (Fig.2c). 
A mixed glassy region
is generated between the true critical 
temperatures, the Curie or N\'eel temperatures in this case, and a
remnant of the clean-limit transition, T$^*$. In this regime, 
perturbations such as small magnetic fields can have dramatic
consequences, because they only need to align the
randomly oriented magnetic moments
of preformed nanosize FM clusters to render the system globally ferromagnetic.
A concomitant percolation induces metallicity in the compounds.
The fragility of the state show in Fig.2c
implies that several perturbations besides magnetic fields should induce
dramatic changes, including  pressure, strain, 
and electric fields \cite{review,book}.
Moreover, the discussion centered on Fig.2, a to c, is independent of the details of
the competing states, and should be valid for the AFI versus 
superconducting (SC) state competition in 
cuprates \cite{alvarez} and many other cases \cite{cooper}.

Calculations that incorporate the effects of phase competition
and quenched disorder have been able to reproduce the huge magnetoresistance
observed experimentally \cite{burgy,burgy2}; this suggests that the CMR effect would
not occur without either competing states or the quenched disorder and interactions
necessary to nucleate clusters. This is
in agreement with experiments for $\rm Re_{0.5}Ba_{0.5}MnO_3$
(where $\rm Re$ is a rare earth element) \cite{tokura-disorder2}, which
can be prepared both in 
ordered and disordered forms for the $\rm Re$-$\rm Ba$ distribution. 
Remarkably, only the latter was found 
to exhibit CMR (Fig.2d). This suggests that when phases compete, the 
effect of (typically small amounts of)
quenched disorder results in dramatic properties that are very
different from those of a slightly impure material \cite{burgy,burgy2,motome,aliaga}. 
Disorder in the regime of phase competition is not a mere perturbation;
it alters qualitatively the properties of the material.

How strong should the disorder be to induce the inhomogeneous patterns discussed here? 
Are there other alternatives?
Studies incorporating long-range effects, such as Coulombic forces \cite{schmalian,schmalian2} or
cooperative oxygen octahedra distortions \cite{burgy2}, suggest that 
very weak disorder, even
infinitesimal disorder \cite{schmalian,schmalian2}, may be sufficient 
to do the job. Calculations without explicit disorder incorporating
strain effects \cite{ahn2}, or within a phenomenological Ginzburg-Landau
theory, also lead to inhomogeneous patterns \cite{milward}. While the discussion
on the details of the origin of the inhomogeneities is still fluid, 
their crucial relevance to understanding the manganites, as originally predicted
by theory \cite{review,book}, is by now widely accepted.
\begin{figure}
\includegraphics[width=3.0cm]{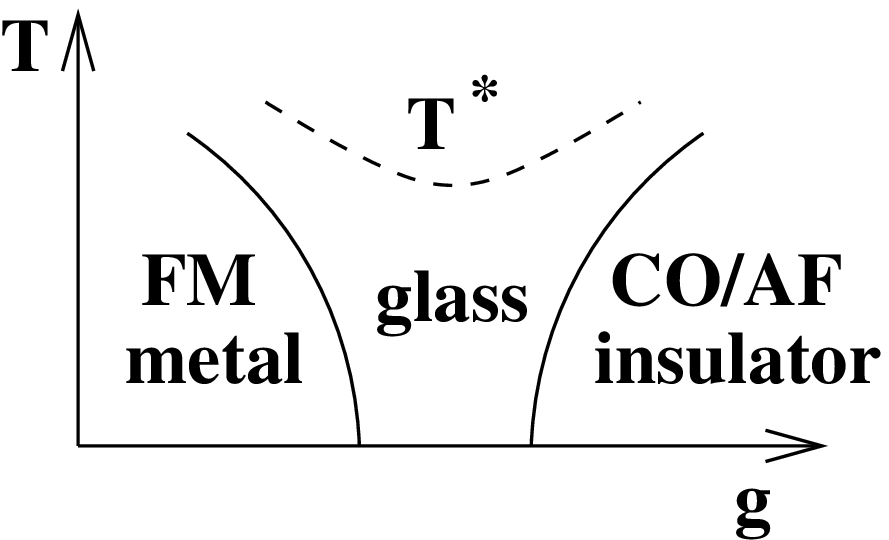}
\includegraphics[width=5.0cm,clip]{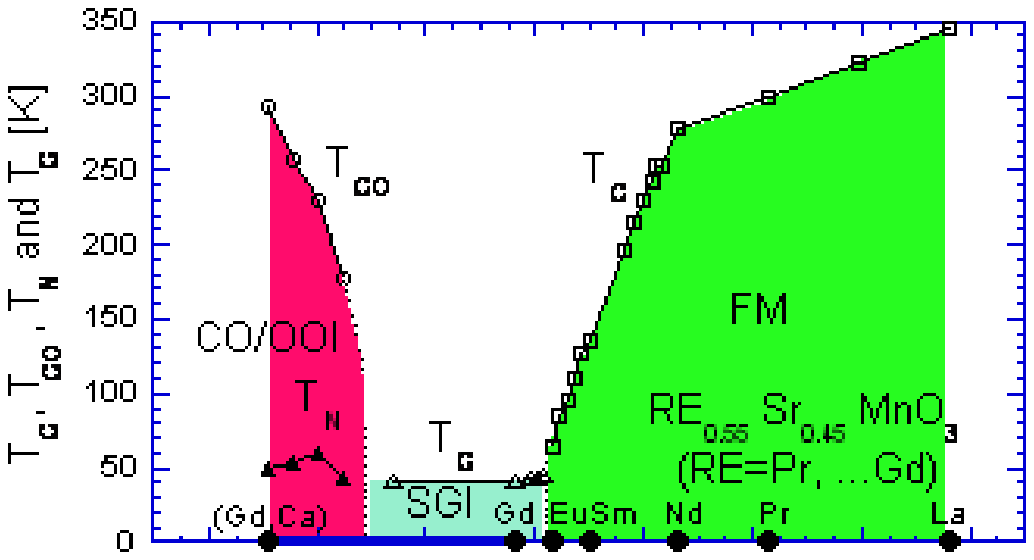}
\includegraphics[width=4.0cm,clip]{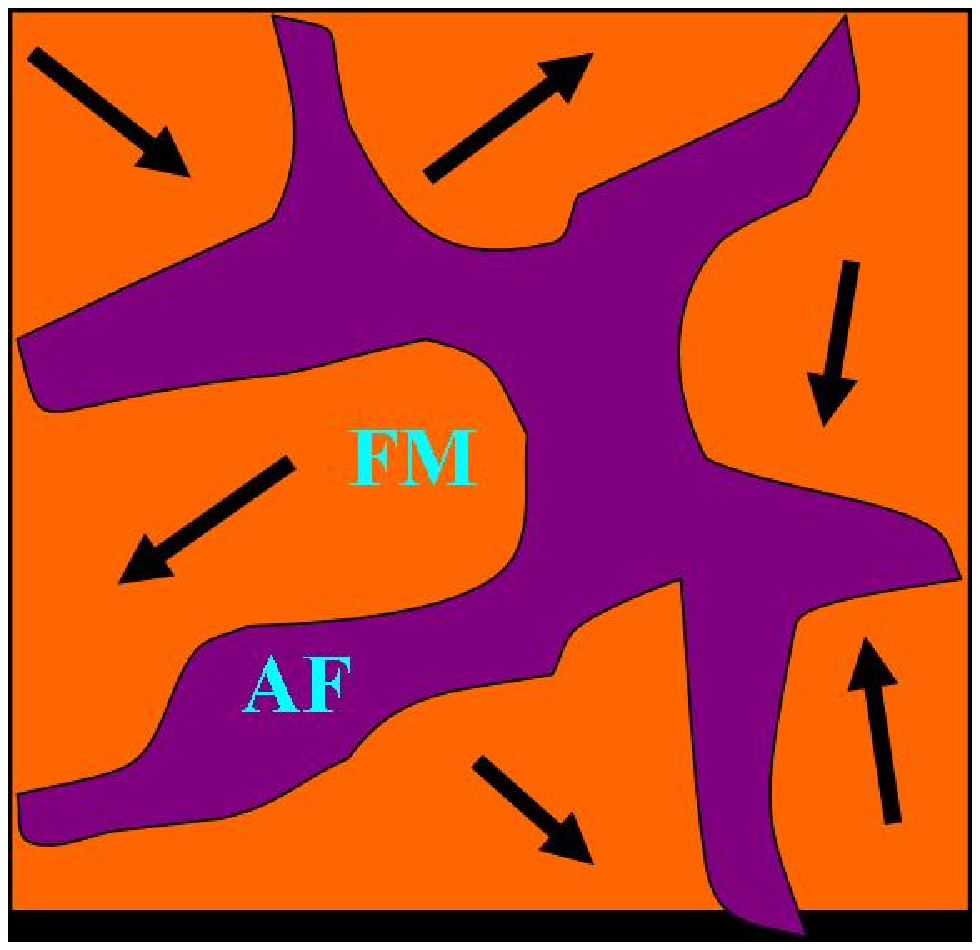}
\includegraphics[width=3.5cm]{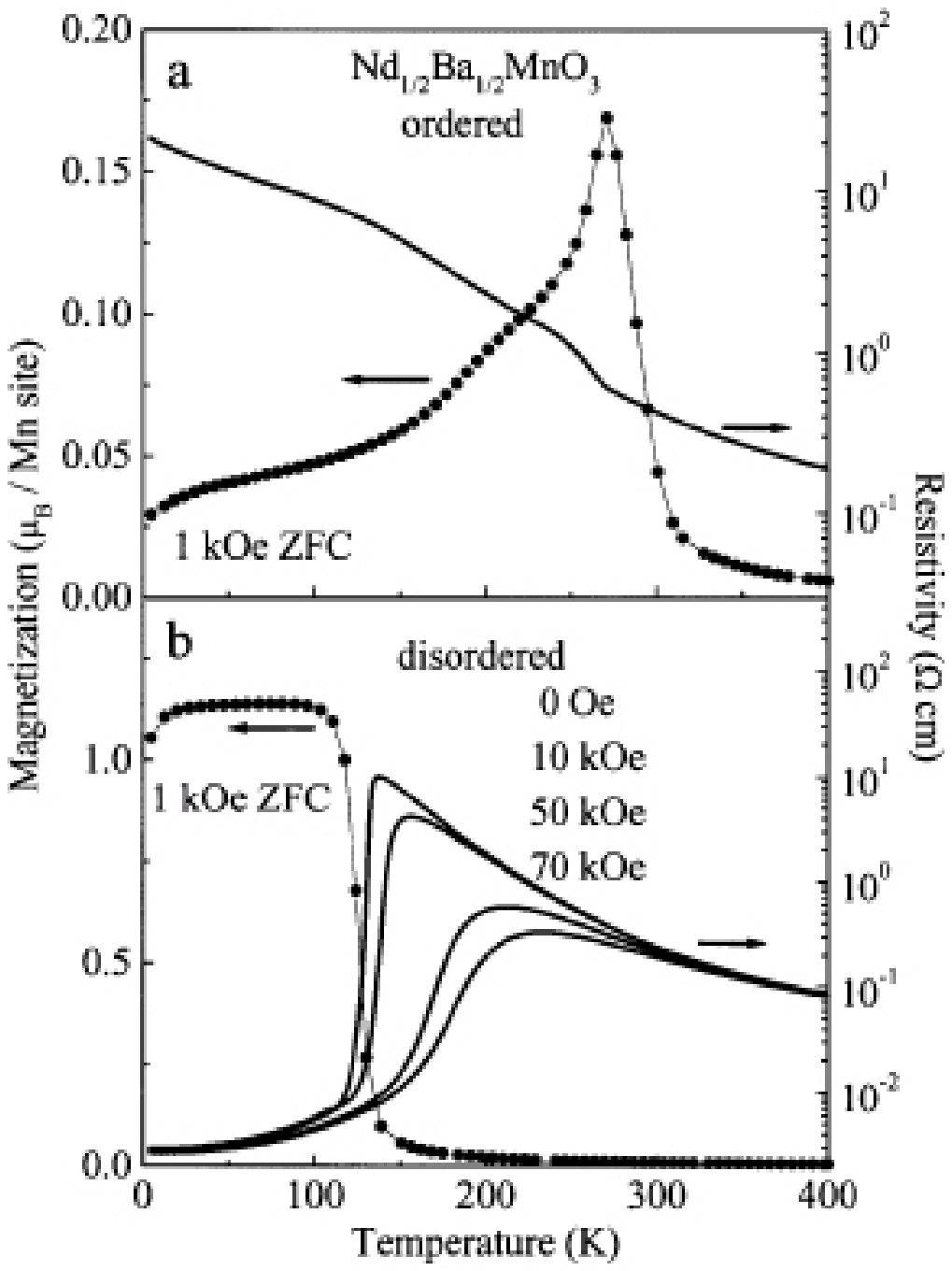}
\caption{\label{fig:manganitas} 
(a) Generic phase diagram of two competing states (here FM metal vs. 
charge-ordered antiferromagnetic (CO/AF) insulator) in the presence of quenched disorder
 \cite{review,book}. $g$ is a generic variable to move 
from one phase to the other (e.g. electronic density or bandwidth).
A glassy mixed-phase state is created and a $T^*$ scale
appears. 
(b) Experimental phase diagram of manganites with large disorder 
\cite{tokura-disorder,tokura-disorder2}. 
Note the disorder-induced 
suppression of the ordering temperatures and the appearance of a 
glass state, as predicted by theory (a). 
Details, and the phase diagram
with weak disorder, can be found in \cite{tokura-disorder}.
(c) Sketch of the proposed CMR state for the manganites,
containing FM clusters with randomly oriented moments, separated by
regions where a competing CO/AF
phase is stabilized \cite{review,book,cheong}.
(d) Resistivity and magnetization versus temperature for the ordered
and disordered structures of $\rm Nd_{0.5}$$\rm Ba_{0.5}$$\rm MnO_3$ 
\cite{tokura-disorder2}. 
Only the disordered crystal has the CMR effect
\cite{tokura-disorder2}.
} 
\end{figure}

{\it Cuprates:}
In the HTSC context, the $\rm La_{2-{\it x}} Sr_{\it x} Cu O_4$ 
(LSCO) phase diagram is usually considered as the universal diagram for cuprates.
However, some investigations 
suggest otherwise \cite{alvarez,sczhang}.
For example, only after Ca is added to $\rm Y Ba_{2} Cu_{3} O_{6+\delta}$ 
(where $\delta$ is the excess of oxygen, and it ranges between 0 and 1)
does its phase diagram
resemble that of LSCO \cite{sanna,sanna2}.
Moreover,  organic superconductors 
do not have a glassy phase between  the
AFI and superconducting states, and they are believed to be cleaner than the 
cuprates \cite{organic2}. This suggests that quenched disorder (or strain, etc.) 
in cuprates may play a role as important as that in the manganites, 
and the exotic underdoped regime and $T^*$ may emerge as a consequence 
of its influence \cite{alvarez}. If so, then it is not sufficient to
consider phase diagrams involving only temperature and hole doping $x$. A
disorder strength axis should be incorporated into the phase diagram
of these materials as well.

Considerable discussion concerning the existence of inhomogeneous
states in cuprates started several years ago when stripes 
were reported in studies carried out with neutron
scattering techniques \cite{tranquada}. These states had been predicted 
theoretically \cite{stripes,zaanen}.
The nontrivial real-space structure of stripes emerges from Hamiltonians 
that do not  break translational invariance, which is a remarkable result. 
However, since
approximations were made in the calculations, it is still controversial
whether stripes do exist in Hubbard 
 Hamiltonians \cite{stripes-theory,bishop,bishop2,sorella}.
Experimentally, the presence of stripes is also a matter 
of debate. Recent
neutron studies of HTSC materials have been interpreted as caused by 
a phase that contains stripes separated by two-leg ladders (Fig.3a) 
\cite{tranquada2,dai}, with spin-gapped properties that could be important for
pairing \cite{ladders}. In addition, doped Ni oxides
and Nd-doped 
LSCOs are widely believed to have stripes \cite{stripes}.

While the stripe debate continues,
scanning tunneling microscopy (STM) investigations
have recently provided additional important information on
the cuprates, unveiling a variety of other inhomogeneous states.
Figure 3b shows a real-space 
distribution of $d$-wave  SC gaps in
$\rm Bi_{2} Sr_{2} Ca Cu_{2} O_{8+\delta}$ (Bi-2212). 
The many colors illustrate the
inhomogeneous nature of the state \cite{davis},
with randomly distributed nanoscale patches. These patterns could be
caused by
phase competition or by a random oxygen distribution.
Other recently synthesized cuprate-based compounds also 
have inhomogeneous states \cite{takagi}, and additionally,
a new charge-ordered ``checkerboard'' state 
has been observed (Fig.3c) \cite{hanaguri}. 
This state also exists in Bi-2212 \cite{yazdani} and appears to
compete with superconductivity.
Understanding these novel states remains a challenge,
but for our purposes 
two issues are crucial: (i)
When scrutinized with powerful 
microscopic techniques, doped HTSC systems reveal inhomogeneous
states. Supporting this statement,
a novel scaling law
for the cuprates was interpreted as produced by a Bardeen, Cooper, and Schrieffer
system in the $dirty$ limit \cite{homes,homes2}.
(ii) The intermediate states between the AFI and SC states
do not seem universal
(they could have stripes, a charge checkerboard, or glassy patterns).
All these characteristics are hallmarks
of complex systems, showing sensitivity to details, as it occurs
in nonlinear chaotic systems.

\begin{figure}
\includegraphics[width=3.5cm]{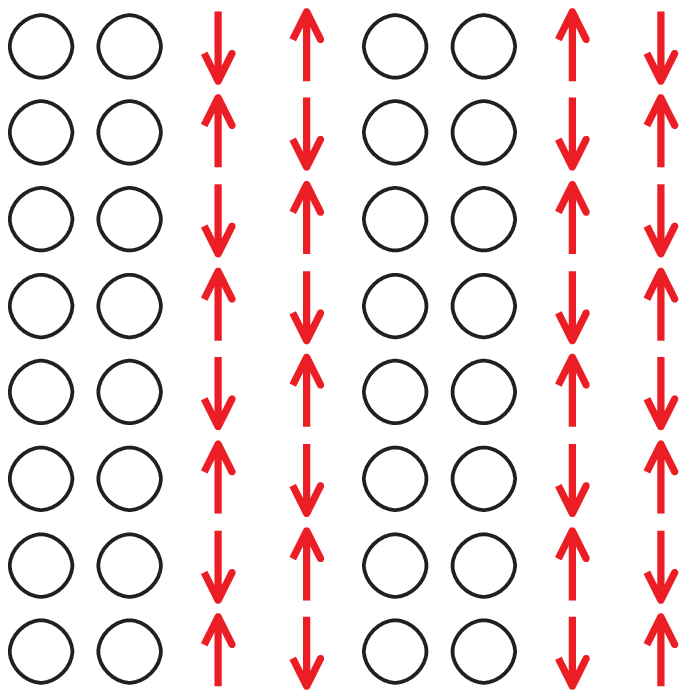}
\includegraphics[width=3.0cm]{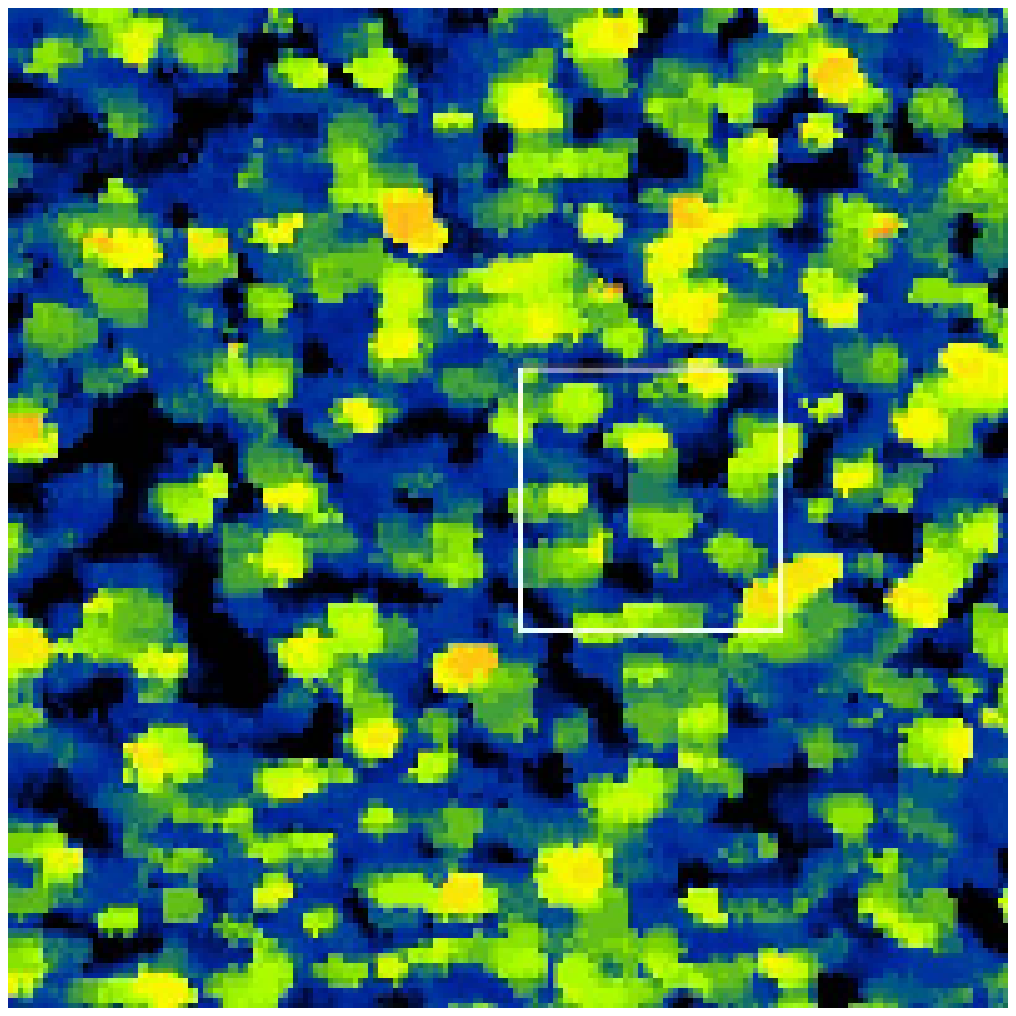}
\includegraphics[width=3.0cm]{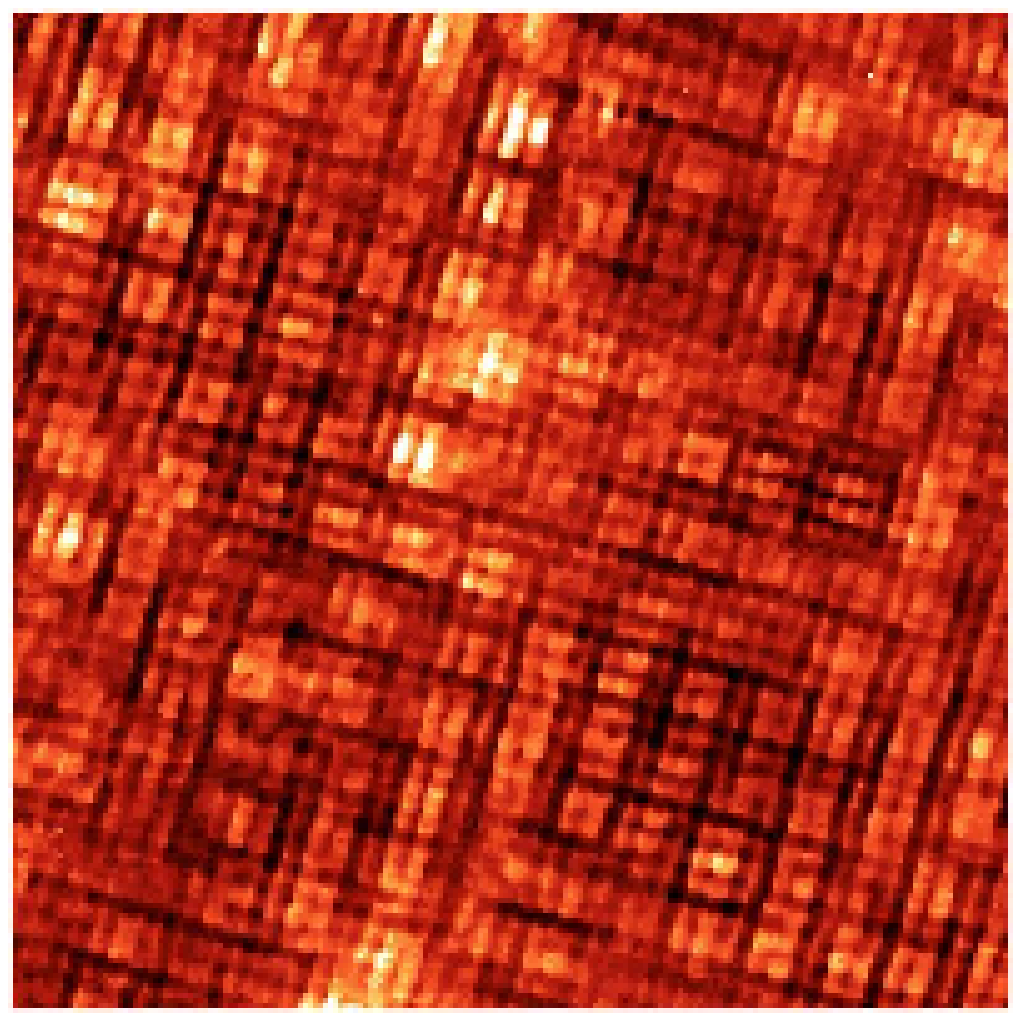}
\caption{\label{fig:patterns} 
Examples of inhomogeneous states in HTSC materials.
(a) Schematic perfect stripes \cite{tranquada2} (circles are holes, 
arrows spins). Real systems may present more dynamical stripes \cite{stripes}. 
(b) $d$-wave SC gap real-space distribution obtained using 
STM techniques \cite{davis}. Inhomogeneities at
the nanoscale are observed (patches). The entire frame is of size
560$\rm \AA$$\times$560$\rm \AA$.
(c) Recently unveiled charge-order state (checkerboard) 
in Na-doped cuprates \cite{hanaguri,yazdani}.} 
\end{figure}

Some additional issues should be remarked upon:
(i) While the most complex behavior in cuprates appears in the underdoped
regime, dynamic electronic
inhomogeneity and competition among the many degrees of freedom
could also underlie the superconductivity even at
optimal doping. 
Are the inhomogeneities and complexity
at the root of the superconducting phase, or are they unrelated?
The discussion continues.
(ii) Interactions can also generate inhomogeneous 
patterns \cite{stripes,kivelson,alvarez},
and the combination of these interactions with
quenched disorder may be at the heart of the complexity in
cuprates. 

Another unexpected property of the cuprates is the giant proximity effect.
This phenomenon has a long story, but recently it has been very carefully
studied using atomically smooth films made  of HTSC compounds
in S-N-S trilayer junctions (S is for superconductor and  N equals normal metal)
\cite{bozovic}. The big paradox here is that
the trilayer behaves as a Josephson junction for N barriers 
a hundred times thicker that the coherence length, $\xi$. Then,
 the normal state can not be featureless,
it must already contain a 
tendency toward superconductivity, which could be in the form
of nanoscale SC islands \cite{alvarez} or phase-fluctuating homogeneous
states \cite{kivelson-emery}.
This proposal leads to an exciting prediction: Under the proper
perturbation, the state with preformed SC clusters 
should  present a gigantic susceptibility toward 
superconductivity \cite{alvarez}. 
This is the analog of CMR in Mn-oxides, 
but translated into Cu-oxide language.
In general, theory predicts that giant responses to external perturbations should 
be far more common than previously anticipated.

\vskip 0.3cm
\noindent{\bf The Case for Complexity in Correlated Electron Systems}
\vskip 0.2cm

\noindent {\it Are TMOs examples of complex matter?}
Considering the general properties of complexity briefly reviewed in the
introduction,
as well as the oxide results discussed in the previous section, it is natural
to wonder whether these systems can be considered as special cases of complex matter.
While complexity is natural when associated with soft 
matter (literally soft, for example polymers and liquid crystals),
it seems out of place in the context of hard materials. But 
the several simultaneously active degrees of freedom  
may conspire to provide a soft 
electronic component to transition metal oxide compounds,
soft not in the physical hardness sense but denoting 
the existence of a multiplicity of nearly degenerate conformations
of the electronic component that can be easily 
modified by external perturbations.
Conventional soft matter is classical ($\hbar$=0), 
but in the electronic systems
described here quantum effects are important.

TMOs are soft in the sense described above as already proposed 
in the HTSC context \cite{kivelson}. They are also
complex, because several effects become simultaneously important and prevent
a simple physical description.
More specifically,
consider one of the popular definitions of complexity recently discussed
in \cite{vicsek}:
``... randomness and determinism are both relevant to the system's overall
behavior. Such [complex] systems 
exist on the {\it edge of chaos} -- they may exhibit almost
regular behavior, but also can change dramatically and stochastically in time 
and/or space as a result of small changes in conditions.'' This definition is
satisfied by manganites, in which a small magnetic field produces 
a drastic change in transport properties, and it may apply 
to underdoped cuprates as well 
 \cite{bozovic,alvarez}. 
When phases compete,  general arguments suggest
that large responses to weak perturbations
should be far more common than previously believed \cite{review,book,alvarez}.
Although the basic rules for electrons 
(i.e., the Hamiltonians) 
are deceptively simple (nearest-neighbor carrier hopping, 
Coulomb or phononic interactions, etc.), the outcomes are
highly nontrivial when phases compete and percolative
physics, as when only a narrow channel exists for electrical conductivity
through a material, is at work.

Another argument can be found
in the known properties of traditional soft condensed matter, which
is a phase of matter between a simple fluid
and a regular solid crystal. In soft matter, large
groups of atoms form regular patterns as in a solid,
but when several of these large groups are considered together
a complex fluid behavior emerges. Typical
examples are polymers: in each large molecule there
is atomic regularity, but an ensemble of them have a
variety of
fluid phases \cite{witten}. This variable behavior is also
present in some TMOs: in manganites,
several experimental investigations have found evidence for 
Jahn-Teller 
ordered small regions (i.e. with a particular form of lattice distortions)
in the state above the FM ordering temperature \cite{review,book}. As a system,
these small Jahn-Teller clusters,
along  with the magnetic clusters present in the same  phase region, 
generate a collective behavior that is 
different than the behavior of the system's individual parts,
and in this temperature range 
colossal magnetoresistance occurs \cite{review,book}. 
Also cuprates may behave as electron liquid crystals, 
intermediate between electron liquid and electron
crystal \cite{kivelson}. Softness in the manganite context
has also been recently discussed \cite{milward}.
Once these concepts are accepted, then the long history of soft-matter
investigations suggests that it is natural to expect new kinds of
organized behavior.
In complex systems, randomness and determinism are simultaneously
relevant, and these are ideas compatible with recent correlated electrons
investigations \cite{burgy,burgy2,motome,aliaga,tokura-disorder}.

Each complex situation in correlated electrons
may lead to a unique state. Some materials may 
have stripes, others may have patches, some
may have phase separation at nanoscales, 
others may have mesoscale phase separation;
the number of states in competition and their nature can lead
to enormous possibilities. This is exciting for applications, but
frustrating for those with a reductionist soul. What is likely is
that new general concepts and paradigms will emerge as guiding
qualitative principles in the study of complex oxides. It will be difficult
to predict the precise shape of the nanopatterns and the phases in competition
unless detailed calculations are performed. But the existence of some
patterns, as well as  giant responses to selected
external perturbations, will be predictable.
Certainly the highest degree of complexity
is expected when many degrees of freedom
 are active simultaneously, and when
many phases with different properties are in competition. 

{\it Theory, phenomenology, computer simulations:}
How can we make further progress in this context?
Investigations involving the
fundamental 
Hubbard and $t$-$J$ 
models are reaching the limits
of our current many-body techniques. 
It appears unlikely
that the large length scales needed to fully capture the complex behavior
of oxides, where percolation is probably very relevant, will be reached
via this path,
and we must focus on the right level of description. As Laughlin and Pines
\cite {laughlin} wrote ``Deduction from microscopics has not explained, and
probably cannot explain as a matter of principle, the wealth of crossover
behavior discovered in the normal state of the underdoped cuprates''.
It is still reasonable that key issues such as the pairing 
mechanism and short distance nature of the dominant states 
can still be analyzed in the context of Hubbard-based approaches,
perhaps supplemented by long-range Coulomb and/or electron-lattice
interactions.
However, the complexity of the resulting states, with 
emerging self-organization and
giant responses, can only be addressed with simpler phenomenological models
that assume competition 
between a few selected states, and analyze its consequences.
For example, the famous linear resistivity and  puzzling 
underdoped behavior of the cuprates, and the
CMR effect in the manganites may only be explainable using 
this coarse-grain approach. 

The logical chain starts with  $ab$ $initio$ calculations
to evaluate the main parameters and couplings, followed by Hubbard modeling 
to obtain the dominant short-distance correlations, and ends with the use
of more phenomenological models \cite{alvarez} 
to handle the long length scales of 
relevance in an electronic complex fluid. The inclusion of both
symmetry and spontaneous symmetry breaking will be important to achieving
these objectives, as will be the inclusion of the effects of disorder
and lattice distortions.
Essential for the success of the present flurry of research in complex systems
is the ability to use high-speed computers to perform unbiased calculations.
By simulating a system made of many small units, the behavior 
of the whole ensemble can be understood
and manipulated much better than with other techniques,
providing new ways of learning and visualizing in this context.

{\it Other systems with similar complex behavior:}
There are many other materials that behave similarly as the TMOs emphasized
in this review. For example, in the area of 
heavy fermions (metals where the effective
mass of electrons is much larger than the bare mass) the presence of ``electronic
Griffiths phases'', inhomogeneous states at zero temperature,
has been described \cite{vlad}, and strong similarities with
the cuprate's phase diagram were unveiled \cite{heavy} (Fig.~1f). In general, 
glassy behavior is expected
near a metal-insulator transition at low temperatures \cite{vlad,vlad2,vlad3}, 
establishing
an interesting connection with the area of investigations known as ``quantum
critical phenomena'' \cite{sachdev}. Glassy dynamics is also observed in
other two-dimensional electronic systems \cite{popovic}. Cobalt oxides \cite{kuhns},
organic materials \cite{organic,organic3} (Fig.~1e), and Ca-doped ruthenates (Fig.~1d)
are other examples. Materials where charge density waves and superconductivity
compete provide other cases of complex behavior \cite{CDW}.
The area of complexity in correlated electrons
is far wider than the two TMOs chosen in this article to focus on.

{\it Complexity in pure states:}
The emphasis of this review has been on self-organization and the complexity
in the electronic sector associated with the existence of several
competing states. This corresponds to the physics of the HTSC cuprates
in the underdoped regime, and the manganites in the CMR regime.
However, complexity in strongly correlated electrons also exists in
the fascinating ground states observed in the clean limit, or far from
the region of phase competition if quenched disorder is present. For example, 
superconducting
ground states   with zero electrical resistance, a Meissner effect, and
unconventional properties ($d$-wave in the cuprates or
spin-triplet pairing in the
ruthenates \cite{maeno}) emerge from simple interactions
among electrons and lattice vibrations. In the manganites, a CE phase exists 
with simultaneous spin, lattice, orbital, and charge order \cite{review}.
The list of exotic phases observed in the clean limit is enormous, and they all
represent emergent phenomena in the sense that their properties cannot be predicted 
easily from the Hamiltonian. The collective behavior of electrons in these
phases is relatively simple, and it  can be described with 
a handful of concepts and parameters. The emergence of simplicity is part of the
complex behavior of electrons \cite{laughlin}.
While in the case of Mn-oxides the inhomogeneities are crucial to understanding
 the CMR effect,
and in cuprates the analogous inhomogeneities are important to rationalize the
curious underdoped regime, they do not provide an obvious mean to comprehend
the origin of all the many exotic ground states.
Thus, with or without quenched disorder, in homogeneous or
inhomogeneous forms, it is clear that systems
of strongly correlated electrons are surprising and the list of their many possible
ground states is far from fully classified. Research producing highly pure samples
is as important as those focusing on the region of inhomogeneities and pattern formations,
leading to complementary insight. Clearly, these compounds are complex in more than one sense.

{\it Applications?}
It is too early to decide if the  
complex properties of correlated oxides could be important for applications,
but several results already provide interesting clues. 
To name a few, the resistance of some oxide films
were unexpectedly found to switch between low and high values upon the application
of voltage
pulses \cite{switch,marcelo}. Also, gigantic magnetoelectric effects
were reported  \cite{kimura}, interfaces of magnetic oxides have been
engineered \cite{yamada}, 
manganites with sharp magnetization steps exist \cite{schiffer},
and manganite nanotubes were prepared \cite{levy-nanotubes,levy3}.
Creating ultra-smooth thin films and artificial superstructures is part
of the avenue toward applications. 
Since complexity appears to be the reason behind the CMR effect, 
complex behavior is conducive
to functionality. 
Relaxor ferroelectrics are also oxides with nanodomains, with
potential applications \cite{relaxor}.
Field-effect transistors  made from TMOs
are another exciting area of research \cite{ahn}:
correlated electron
 materials could present phase transitions in the presence of
electric fields  since these fields can alter the carrier concentration. 
It is the diversity of behavior,
namely the many possible metallic, insulating, magnetic, 
superconducting, and ferroelectric phases 
of strongly correlated systems, that makes these types of investigations
so exciting.

\vskip 0.3cm
\noindent{\bf Conclusions}
\vskip 0.2cm

\noindent TMOs are certainly not as simple as standard metals.
The many active degrees of freedom -- spin, charge, lattice, orbital --
interact in a nonlinear, synergetic manner, leading to an intrinsic complexity.
STM, neutron and X-ray scattering, and other microscopic techniques,
are crucial to unveiling the subtle 
nanoscale phase separation tendencies that induce a variety of
real-space patterns.
Charge transport in oxides is quite different from the
free flow in simple metals: an isolated charge 
strongly perturbs its environment, inducing a polaron, which often
attracts other polarons to form larger structures. 
To capture this physics, it is important to incorporate several 
ingredients, including powerful nonperturbative many-body techniques,
phenomenological approaches, and the effects of lattice
distortions, strain, and quenched disorder. All these ingredients appear
equally important.
Phase competition rules the behavior of these compounds:
while the energies characterizing each phase (such as gaps) 
can be fairly large, 
at particular carrier densities or bandwidths
the energetic proximity of two phases 
introduces a lower hidden energy scale, and small perturbations
cause huge  responses, not via the melting of the state analyzed
but by its replacement by a very different one.

Establishing electronic complexity in hard materials
as a fundamental area of research
will create scientific 
relations with other popular fields
of investigations. For instance, the existence of
complexity in biological systems is clear, and analogies between
proteins and spin glasses, both of which have a
distribution of barrier heights among competing 
nearly degenerate states, have often been remarked upon \cite{wolynes}.
In fact, most correlated electronic systems
exhibit exotic glassy behavior with notoriously slow dynamics \cite{schiffer2,levy2},
establishing one of the prime connections between traditional biological or soft systems
and the complex states described here.
Biological physics is one of the major
frontiers for physics in the new millennium and complexity certainly
arises in macromolecules and
complex fluids. A common language can also be established with other
broad fields: for instance, in nuclear matter the self-generation of
structures is under much discussion as well \cite{pieka}.

A novel paradigm
involving ``complexity in correlated electron 
materials'' will help to focus on the right
level of description, on the expected emergence of patterns, and on separating
the physics of the individual phases from properties that arise
from phase competition.
Controlling  the spontaneous tendencies toward
complex pattern formation 
may open the way to achieving emergent functionalities in correlated electrons
systems. The enormous diversity of phases in oxides provides a wide
range of combinations to explore. 
Complexity and functionality are rapidly developing into
the most exciting frontiers in the active area of
strongly correlated electrons.



\end{document}